\documentclass[amssymb,aps,pre,twocolumn,showpacs,superscriptaddress,groupedaddress]{revtex4-1}  % for review and submission
\usepackage{graphicx}  % needed for figures
\usepackage{dcolumn}   % needed for some tables
\usepackage{bm}        % for math
\usepackage{amssymb}   % for math
\usepackage{mathptmx}
\usepackage{color}

\DeclareMathAlphabet{\mabold}{OT1}{cmr}{bx}{it}

\usepackage[inline]{trackchanges}
\addeditor{LK}
\addeditor{ND}
\addeditor{ND1}

\begin{document}

\title{Instability of fluid films on a thermally conductive substrate}

\author{N. Dong and L. Kondic}
\affiliation{Department of Mathematical Sciences, New Jersey Institute of Technology, Newark, NJ 07102 USA}

\date{\today}

\begin{abstract}
We consider thin fluid films placed on thermally conductive substrates and  exposed to time-dependent spatially
uniform heat source.  The evolution of the films is considered within the long-wave framework in the regime such that 
both fluid/substrate interaction, modeled via disjoining pressure, and Marangoni forces, are relevant.  We analyze
the problem by the means of linear stability analysis as well as by time-dependent nonlinear simulations.  The main 
finding is that when self-consistent computation of the temperature field is performed, a complex interplay
of different instability mechanisms results.   This includes  either monotonous or oscillatory dynamics of the 
free surface.  This oscillatory behavior is absent if the film temperature is assumed to be slaved to the current 
value of the film thickness.   The results are discussed within the context of liquid metal films, but are of relevance to dynamics
of any thin film involving variable temperature of the free surface, such that the temperature and the film interface itself evolve on
comparable time scales. 
\end{abstract}

\maketitle

\section{Introduction}

Instabilities of thin fluid films are relevant in a variety of different contexts, with 
many of these involving temperature variations  that lead to modified  
material properties.   In particular, the surface tension of many 
liquids is sensitive to temperature,  resulting in well known Marangoni 
effect, that has been discussed 
 in excellent review articles~\cite{oron_rmp97,cm_rmp09} and books~\cite{COLINET}.

Instabilities due to Marangoni effect have been studied extensively,
and we will focus here exclusively on the settings that involve deformation of the 
free surface.  The studies are often
carried out using the long-wave approach; within this framework, a significant 
body of work has been established in the recent years, including extensive research 
on linear and weakly nonlinear instability mechanisms~\cite{podolny05,morozov15,nepom_jfm15}, as well as discussion of 
monotone and oscillatory type of Marangoni effect governed instabilities~\cite{COLINET,shklyaev10,shklyaev12,samoilova14}
(only a subset of relevant works is listed here).
While most of the works have focused
on the regime where gravitational effects are relevant, there is also an increasing
body of work considering the interplay between the instabilities caused by Marangoni effect and 
by liquid-solid interaction that becomes important for the films on 
nanoscale, see, e.g.,~\cite{warner02,atena09,khenner_pof11,trice_prl08}.    
Understanding the influence of Marangoni effect on 
film stability  is simplified in the settings where temperature of the film surface
could be related in some simple way to its thickness; however it is not always
clear that a simple functional relation can be accurately established, particularly
in the setups such that the temperature field and the film thickness
evolve on the comparable time scales so that the temperature of the fluid 
may be history dependent.  

One context where thermal effects are relevant involves metal 
films on nanoscale thickness exposed to laser irradiation.   The energy provided
by laser pulses melts the films, and, while in the liquid state, these films evolve on a 
time scale that  is often comparable to the pulse duration (tens of nanoseconds).   The flow
of thermal energy during this short time leads to a complex setup that 
involves heat flow not only in the metal film but also in the substrate, phase
change (both melting and solidification), possible ablation, and chemical effects. 
Coupling of these effects to fluid dynamical aspects of the problem is just 
beginning to be understood~\cite{ajaev_pof03,trice_prb07,trice_prl08, khenner_pof11,langmuir15}.  

This paper focuses on fundamental mechanisms involved
in the influence of thermal dependence of surface tension for
films evolving on thermally conducting substrates, and therefore
considers only the basic aspects of the problem, ignoring the effecs
of melting/solidification, ablation, or temperature dependence of 
other material properties.     For definitiveness, we use the material 
parameters appropriate for liquid metals.   
The substrate is considered to be thermally conductive, but otherwise 
uniform.     Since the motivation
comes from nanoscale films, we do not include gravity, but we do consider 
substrate/film interaction via disjoining pressure model that  allows for natural definition of a contact 
angle.  It should be also noted that inclusion of fluid/solid interaction  is necessary if one wants to 
consider film instability on nanoscale (without its inclusion, an isothermal film never 
breaks down, contrary to experimental findings).  While, as mentioned above, a 
significant body of work considering the influence of Marangoni forces on thin film stability
has been established, we are not aware of any work considering the interplay of Marangoni 
effect and fluid/solid interaction by fully self-consistent computation of the thin film evolution, 
and the temperature field, in fully nonlinear regime.   

The rest of this papers is organized as follows.  We formulate the model in Sec.~\ref{sec:formulation}.
Section~\ref{sec:results} discusses the influence of Marangoni effect for a film of fixed (time-independent)
thickness in Sec.~\ref{sec:fixed}, and then in Sec.~\ref{sec:evolve} for an evolving film.   In
Sec.~\ref{sec:large} we remove the constraint of small domain size and consider large domains
that allow for mode interaction in both two and three spatial dimensions (2D and 3D).   Section~\ref{sec:conclusions}
is devoted to the conclusions.   The parameters used as well as derivation of the models used  are 
given in the Appendix.

\section{Model formulation}
\label{sec:formulation}

We start by discussing in Sec.~\ref{sec:marangoni} in rather general terms inclusion of Marangoni effect in the long-wave model. 
Then, in Sec.~\ref{sec:cond} we focus on discussing temperature computation, and the coupling between
the evolutions of temperature, and of film thickness itself.  We will see that proper accounting for film evolution when 
computing the temperature may be crucial for understanding the influence of Marangoni effect on film stability.

\subsection{Thin film with Marangoni effect}
\label{sec:marangoni}

We will analyze the influence of Marangoni effect within the long-wave framework that allows to 
obtain an insight into the most important aspects of the problems and carry out 
simulations at modest computational cost.   The price to pay is approximate
nature of the results, in particular in the context of liquid metal films that 
are characterized by large contact angles and fast evolution that
suggests that inertial effects (not included in the standard version of the long-wave
framework considered here) may be relevant.    However, despite the fact
that all the assumptions involved in deriving long-wave approach are not strictly satisfied, 
one can obtain reasonably
accurate results when using the long-wave approach to explain physical experiments - see, 
e.g.,~\cite{kd_pre09,fowlkes_nano11,lang13,NanoLett14},
or even when comparing to direct numerical solvers of
Navier-Stokes equations~\cite{mahady_13}.   

Within the long-wave framework, one reduces the complicated problem of 
evolving free surface film into a single 4th order nonlinear partial different equation of
diffusion type for the film thickness, $h$, that expresses conservation of mass of 
incompressible film  and reads  ${\partial h / \partial t} + \nabla \cdot (h{\bf v}) =0 $, where
${\bf v}$ is the fluid velocity, averaged over the film thickness.   This velocity can be related to 
the pressure gradient.  To model Marangoni effect, it is typically assumed that surface
tension, $\gamma$,  is a linear function of temperature: $\gamma(T) = \gamma_0 + \gamma_T T$,
where $\gamma_0 = \gamma(T_0)$, and $\gamma_T$ is (for most of the materials) a negative constant.  
In the present work, $T$ is defined relative to some reference temperature, $T_0$ (we will use 
room temperature), and non-dimensionalized as described below.   In non-dimensional form,
the evolution equation is 
as follows
\begin{equation}\label{eq:evol}
 % \begin{split}
   \frac{\partial h}{\partial t} + \nabla \cdot (h^3 \nabla \nabla^2 h)  + K \nabla \cdot [h^3  f'(h) \nabla h] 
   + D \nabla \cdot (h^2 \nabla T) = 0~.
 % \end{split}
  \end{equation} 
Here,  $\nabla = (\partial /\partial_x), \partial / \partial y)$, and $(x,y)$ are the in-plane
coordinates.   The second term is due to surface tension (with pressure proportional to the film curvature 
that is approximated by $\nabla^2 h$), and the remaining two terms are due to 
solid/fluid interaction and Marangoni effect, respectively.  The function $f(h)$, proportional to disjoining
pressure, is assumed to be of the form   
\[
f(h) =  (h_*/h)^n - (h_*/h)^m\, ,
\]
where we use $(n,m) = (3,2)$ as motivated by direct comparison to the
experimental results for Cu films~\cite{lang13}.     
Next,  we define $t_s = 3\mu l_s/\gamma_0$ as the time scale, where $l_s$ is a chosen length-scale (we use typical 
film thickness of $10$ nm).     The non-dimensional parameters are then specified by
$K=\kappa l_s/\gamma_0,$ $D = 3\gamma_T/(2 \gamma_0)$, and $\kappa$
is related to Hamaker' s constant, $A$, by $A = 6 \pi \kappa h_*^3l_s^3$.    
The reader is referred to Appendix~\ref{sec:param}
for the values of the material parameters used, to~\cite{dk_pof07} for extensive
discussion regarding inclusion of disjoining pressure in the long-wave framework, 
to~\cite{trice_prb07,kd_pre09,lang13} for the use of the long-wave  in the context of modeling liquid metal films, and 
to~\cite{oron_rmp97,cm_rmp09,COLINET} for the discussion of Marangoni effects in a variety of settings.
The numerical solutions of Eq.~(\ref{eq:evol}), discussed in what follows, are obtained using the spatial discretization
and temporal evolution as described in e.g.~\cite{DK_jcp02}, with the grid size equal to $h_*$; such discretization is sufficient 
to ensure accuracy.

\subsection{Thin film on a thermally conductive substrate}
\label{sec:cond}

So far, the presentation applies to any situation where temperature gradients are present.   Let us now focus on 
the setup of interest here, and that is a film exposed to an external heat source (such as a laser for experiments
done with metal films), and is placed on a thermally conductive substrate, such 
as SiO$_2$.   To start, consider a spatially uniform film, exposed to an energy source, and in formulating the model 
describing the temperature of the film, ignore convective effects, and furthermore consider only 
the heat flow in the $z$ direction, normal to the plane of the film.
Then, the temperature of the film (and of the 
substrate) can be modeled by diffusion equations (with a source term)  for the film and for the substrate, coupled by 
appropriate boundary conditions
\begin{equation}
{\partial T_i \over \partial t} = K_i  {\partial^2 T_i \over \partial z^2}  + Q_i,\quad (i = 1,2)\ ,
\label{eq:heat}
\end{equation}
where $i  = 1,~2$ stands for the film and for the substrate phase, respectively.   The parameters entering the equation are
listed in Appendix~\ref{sec:heat}, where we also define the temperature scale that is used throughout;  here we only note that the source term, 
$Q_1$, also includes absorption of heat in the film, and is of the functional form 
\begin{equation}
Q_1 = C F(t)\exp(-\bar \alpha (z - h))\, , 
\label{eq:q1}
\end{equation}
where $C$ is a constant determined by the intensity of the source (laser), and $\alpha$ is the (scaled) coefficient of absorption.  
We will assume that the substrate does not absorb heat ($Q_2 = 0$); this is 
appropriate for SiO$_2$ that is transparent to radiation.  In fluid modeling that follows, we will also assume that the 
substrate remains solid.
The boundary conditions include no heat transfer at the free surface; in the spirit
of the long-wave approach this simplifies to $\partial T(z)/\partial z|_{z=h} = 0$ even for nonuniform films; at $z=0$ 
we use continuity of temperatures and heat fluxes, therefore ignoring thermal resistance there, and at the bottom 
of the substrate, we put $T(-h_s) = 0$ (room temperature).  Ignoring heat flow in the in-plane direction can be justified
by relatively slow time scale of heat conduction in the substrate (due to low heat conductivity of SiO$_2$).   
Further studies of the importance of the in-plane heat transfer would be however appropriate and should be considered
in future work.   In the present work, we focus only on the main aspects of the connection between heat conduction and
film evolution.  We note that similar approach (of considering heat transfer in the out-of-plane direction only)
 has been used in existing studies, see, e.g.,~\cite{trice_prb07, fowlkes_nano11}.

\begin{figure*}[t!]
\includegraphics[width= 0.33 \textwidth]{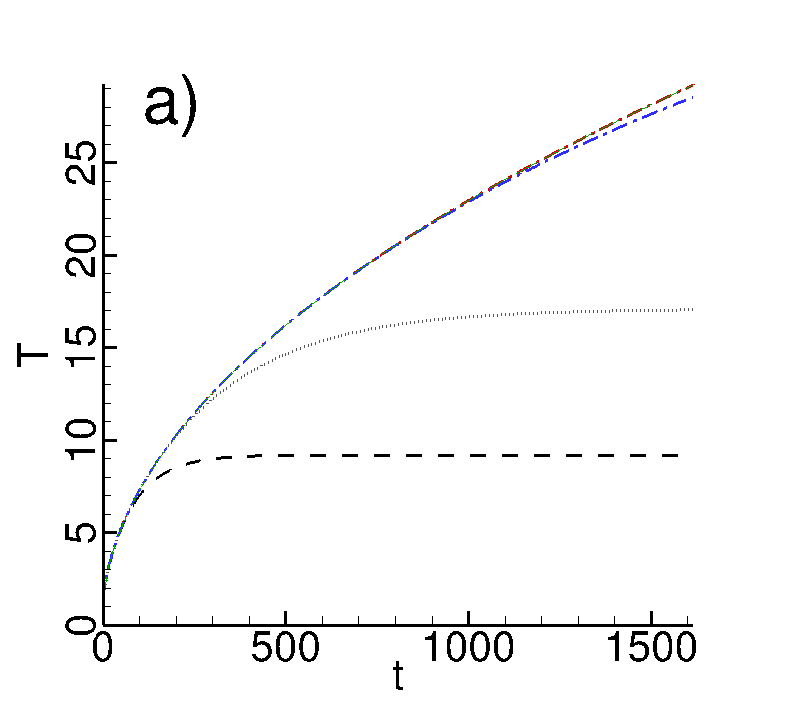}
\includegraphics[width= 0.33 \textwidth]{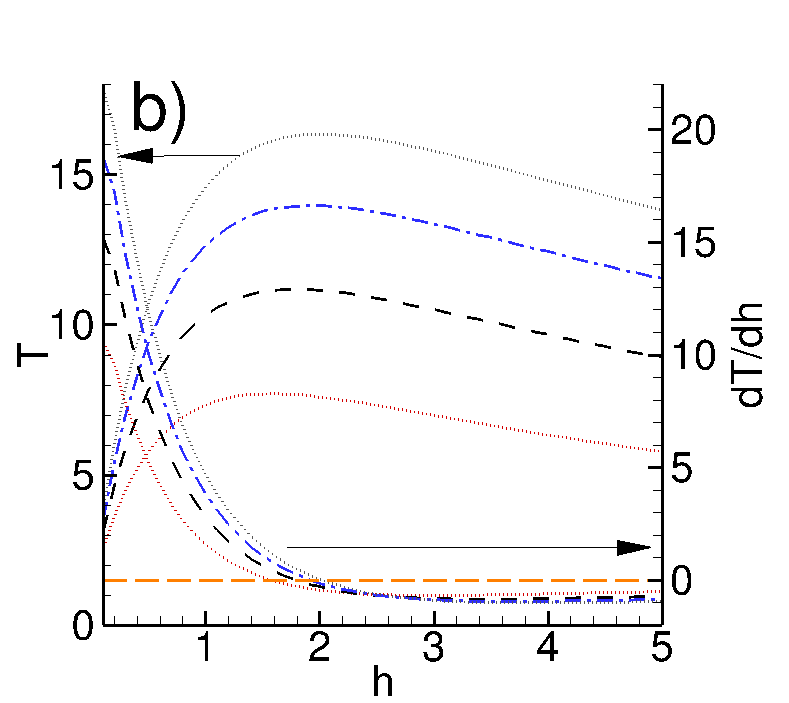}
\includegraphics[width= 0.33 \textwidth]{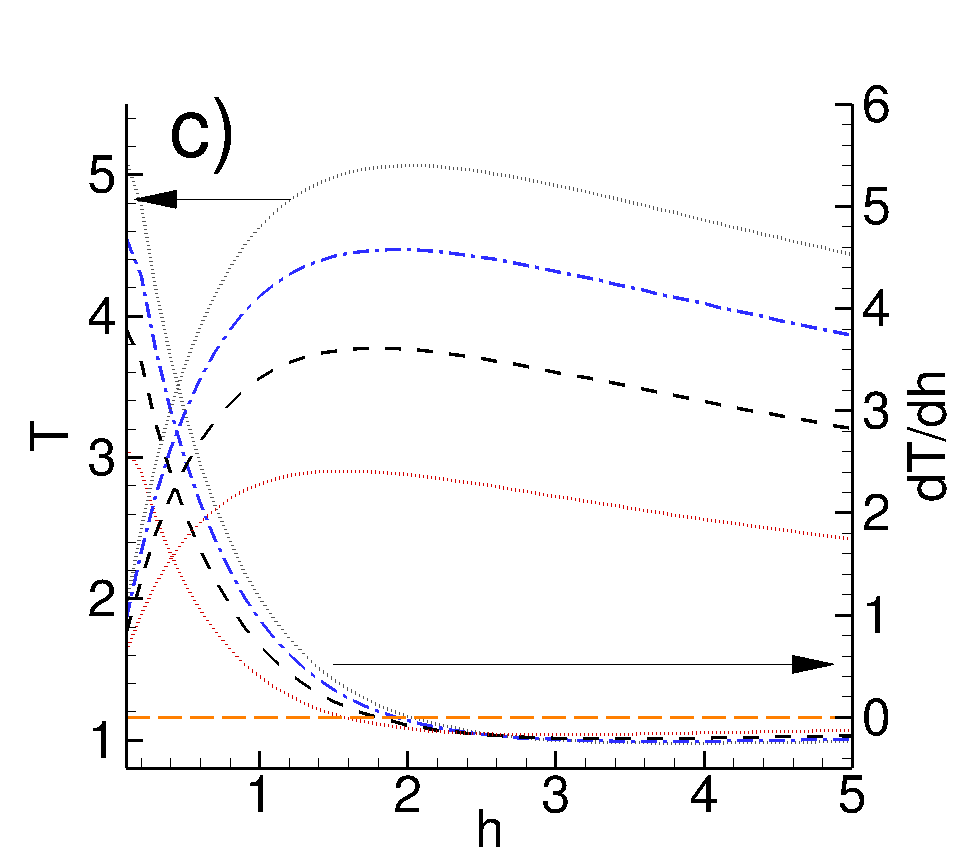}
\caption{\label{fig:temp_uni} (Color online)
a) Temperature evolution of fixed thickness film under uniform laser pulse computed
analytically assuming infinite  $\mathrm{SiO_2}$ thickness  $h_s$ (red dash-dotted),  and numerically
using  $h_s=10$ (black long dashed), $h_s=20$ (grey dotted), $h_s=50$ (blue dash-dotted), and 
$h_s=100$ (green dashed). (Note that the numerical result with $h_s = 100$ overlaps the 
analytical one.)  In this and all the following figures the material and laser properties are from Table~1 in 
Appendix~\ref{sec:param} if not specified differently; in particular here laser energy density is given by $E_0$.  
(b-c)
Temperature and $\partial T/\partial h$ (with $t$ fixed) of the free surface obtained by solving numerically Eq.~(\ref{eq:heat}) assuming
fixed film thickness, $h$.  Here,
$t = 100$ (red  dotted), $t = 200$ (black dashed), $t = 300$ (blue dash dotted), $t = 400$ (grey dotted); the arrows 
indicate the axis related to the set of curves.   Orange long dashed line indicates $\partial T/\partial h=0$.
We use time-independent source term here, $F(T) = const$  in Eq.~(\ref{eq:q1});  
laser energy density is $E_0$ (b) and $E_0/4$ (c).    We show both parts (b) and (c) (that 
are visually similar, modulo different scales) for later reference.
}
\end{figure*}

\section{Results}
 \label{sec:results}
 
We first consider in Sec.~\ref{sec:fixed} a film of fixed thickness, and discuss via linear stability analysis the influence
that Marangoni effect has on film stability in such a setup.   The temperature is here calculated either by using the analytical
solution, discussed in Appendix~\ref{sec:heat}, or by directly solving Eqs.~(\ref{eq:heat}).  We will see
that the analytical solution gives good approximation of the temperature field as 
long as the substrate is sufficiently thick (in the considered setup characterized by a fixed film thickness).    
Then, we proceed in Sec.~\ref{sec:evolve} by discussing the setup where both the
temperature and the film thickness evolve on comparable time scales, and show that inclusion of film thickness evolution in 
the formulation modifies strongly the influence of Marangoni effect on film stability: the temperature of the film is influenced
considerably by the history of the film evolution.    Section~\ref{sec:large} then considers the influence of Marangoni effect on 
film stability in large domains, both in 2D and in 3D.  The parameters that are used are as given in 
Table~1 in Appendix~\ref{sec:param}, except if specified differently. 

\subsection{Marangoni effect for a film of fixed thickness}
\label{sec:fixed}

Equations~(\ref{eq:heat}) are solved using standard finite difference method, with spatial 
derivatives discretized using central differences and Crank-Nicolson method implemented for temporal evolution; 
we use $160$ grid points for each of the domains (film, substrate) - this value is sufficient to ensure convergence.
Not surprisingly, the numerical solutions show that the temperature of the film is essentially  $z$-independent, as 
also discussed in~\cite{trice_prb07}.  
Therefore, for simplicity of notation we will from now on 
assume that $T = T(h,t)$.   
The outlined thermal problem, for fixed (time-independent) $h$ and in the limit of infinite substrate thickness, 
$h_s \rightarrow \infty$, also
allows for a closed form solution for $T(h,t)$; see~Appendix~\ref{sec:analytical} 
for a derivation (we will refer to this solution as the analytical one).   Figure~\ref{fig:temp_uni}(a) compares the analytical solution with the
numerical one.  We see that for large $h_s$ there is an excellent agreement between the two, 
as expected.  For smaller values of $h_s$, the numerically computed temperature saturates 
due to the boundary condition at $z = -h_s$.   
Figure~\ref{fig:temp_uni}(b) - (c) shows $T$ and $(\partial T/\partial h)|_t$ (we omit subscript $t$ for simplicity from now on), 
as a function of $h$.  
The main feature of the solution is that $T$ is a non-monotonous function of $h$; an
intuitive explanation is that for thin films, only limited amount of energy gets absorbed, and the temperature remains low; 
for very thick films, the temperature remains low due to a large mass of the fluid that needs to be heated; as an outcome, there is 
a critical thickness at which $T$ reaches a maximum value.    We present results for two source energy densities that we will reference 
later in the text.

The next step is to couple the thermal problem with the fluid one and use the $T$ resulting from Eq.~(\ref{eq:heat}) in Eq.~(\ref{eq:evol}).  
For simplicity, we will limit the consideration to two spatial dimensions so that $h = h(x,t)$ in Eq.~(\ref{eq:evol}).
An initial insight can be reached by carrying out linear stability analysis (LSA)  of a base state of flat film of thickness $h_0$ perturbed
as follows: $h = h_0 (1 + \epsilon \exp(iqx + \sigma t))$.   Assuming that $T$ is a linear function of $h$ (we discuss this
further below), with 
${\partial T / \partial h} = G = {\rm const}$, 
one finds the dispersion relation 
\[
\sigma (q) = h^3_0 q^2 (-q^2 + P_0)\, ,
\]
where
\[
P_0 = K f'(h_0) + D_1/h_0\, ,
\]
and  
\[
D_1 =  {(3 \gamma_T/ (2\gamma)} G\, .
\]
Then, for $P_0>0$,
the most unstable wavelength, $\lambda_m$, and the corresponding growth rate, $\sigma_m$, are
\begin{equation}\label{eq:LSA}
\lambda_m = {2 \pi \over \sqrt{P_0/2}},\quad\quad \sigma_m = {h_0^3 P_0^2\over 4}\ .
\end{equation}
The LSA predicts exponential decay of any perturbation for the films such that  $P_0<0$. 
Considering the films of dimensionless thickness $h_0 \approx 1$, we see from Fig.~\ref{fig:temp_uni}(b) - (c)  that $G>0$ and therefore
Marangoni effect is stabilizing.  
For thicker  films, the LSA predicts increased instability; however note that for such films $|G|$ is rather small (for the present choice of parameters), and 
destabilizing effect of disjoining pressure is very weak, so that evolution is expected to proceed with small growth rate, suggesting
that instability could occur only on very long time scales.

 \subsection{Marangoni effect for an evolving film}
 \label{sec:evolve}
 
The analytical solution for temperature, plotted in Fig.~\ref{fig:temp_uni} and 
discussed in Appendix~\ref{sec:analytical}, as well as the numerical solutions shown in Fig.~\ref{fig:temp_uni} assume
that the film itself does not involve.   However, since thermal and fluid problem are coupled, and furthermore since they evolve on 
comparable time scales (as it will become obvious from the following results, or based on simple dimensional arguments for
the time scale governing the  heat flow compared to the inverse of the growth rate for film instability), 
it is not clear that this assumption is appropriate, 
and it is also not obvious what is its influence on the results.   To answer these questions, we will next consider fully coupled problem, 
where we solve numerically Eq.~(\ref{eq:evol}), while self-consistently computing the temperature by solving the system of diffusion Eqs.~(\ref{eq:heat}).  
We will first consider uniform source term, and then a Gaussian one.   The initial condition is a film perturbed 
by a single cosine-like perturbation of the wavelength corresponding to $\lambda_m$ obtained from the LSA with Marangoni 
effect excluded.    The initial temperature (at $t = 0$) of the film 
and the substrate is taken to be the room temperature, so $T(t=0) = 0$.  The boundary conditions for the thin film equation~(\ref{eq:evol}) are of
no-flux type, with the first and third derivative vanishing at the domain boundaries.

\begin{figure}[t!]
\includegraphics[width= 0.48 \textwidth]{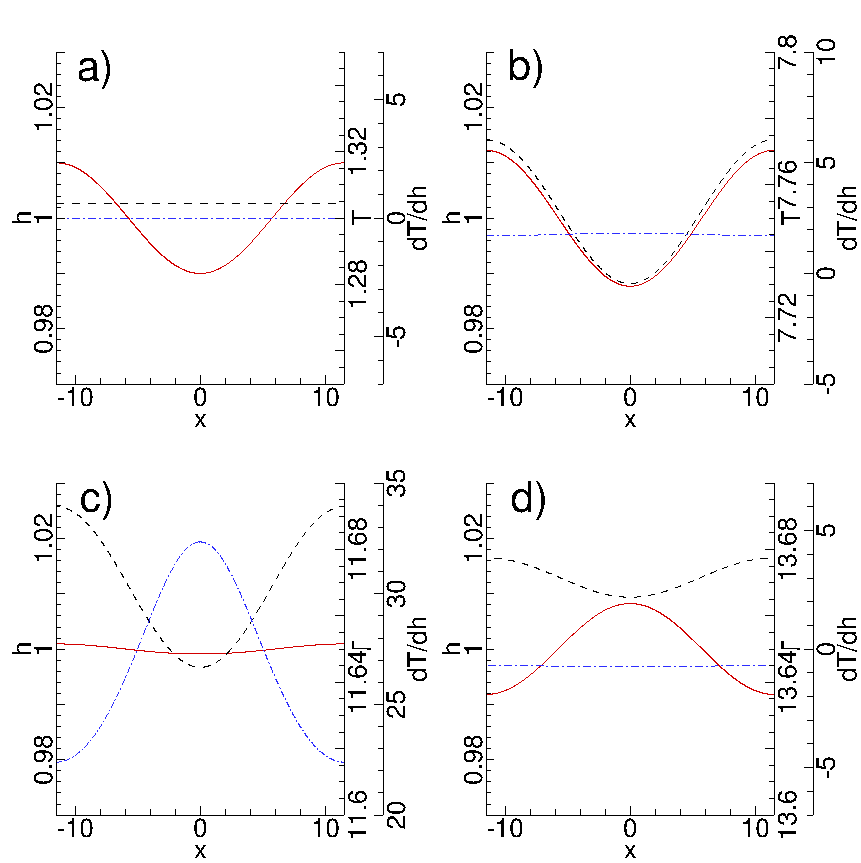}
\caption{\label{fig:profiles_uni}
(Color online) Evolution of the film thickness, $h$ (red solid), temperature, $T$ (black dashed), and temperature gradient, 
$\partial T/\partial h$ (blue dash dotted) as a result of self-consistent time-dependent computations of the film thickness and temperature. 
The domain size is $\lambda_m$ defined by Eq. (\ref{eq:LSA}) without Marangoni effect ($D_1=0$).
The times shown are: $t=0$ (a), $t=113$ (b), $t=258$ (c), $t=355$ (d). 
}
\end{figure}

Figure~\ref{fig:profiles_uni} shows few snapshots of $h$, $T$, and $\partial T/\partial h$.   
Initially, (a) $h$ is perturbed, and $T$ is constant.   The perturbation in $h$ grows (b) due to destabilizing disjoining pressure, and leads to a perturbation in 
$T$.  This perturbation stabilizes the film (the fluid flows from hot to cold), leading to essentially flat $h$, but $T$ and $\partial T/\partial h$ are delayed and are
not uniform (c).   This nonuniform temperature  induces further evolution of the film 
profile and `inverted' perturbation (d), that is again stabilized by Marangoni flow.   This process continues, leading to 
damped oscillatory evolution of the film.   

\begin{figure}[t!]
\includegraphics[width= 0.52 \textwidth]{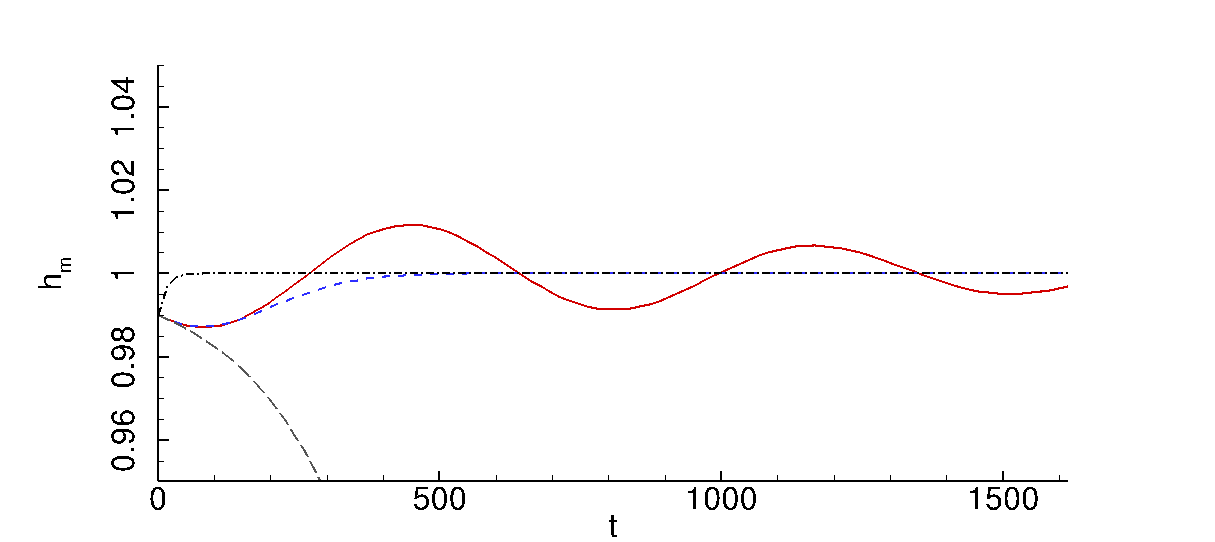}
\caption{\label{fig:h_evol_uni}
The film thickness, $h_m$, at $x_m = \lambda_m/2$ using different approaches to compute the temperature:  
self-consistent time-dependent temperature computation (red solid), the analytical solution of
Eq.~(\ref{eq:heat}) assuming fixed film thickness (blue dashed), and using fixed $G = 3.0$ obtained
from Fig.~\ref{fig:temp_uni}(b) (black dash dotted).
The evolution computed by ignoring Marangoni effect all together is shown as well (grey long dashed).
}
\end{figure}

Figure~\ref{fig:h_evol_uni}  shows the film thickness at the middle of the domain, $h_m = h(x_m = \lambda_m/2)$ as 
a function of time  for few different approaches used to  compute the film temperature: (i) 
the self-consistent time-dependent solution of Eq.~(\ref{eq:heat}) coupled with Eq.~(\ref{eq:evol}) (the same approach
used to obtain the results shown in Fig.~\ref{fig:profiles_uni}); 
(ii) the analytical solution of Eq.~(\ref{eq:heat}) assuming fixed film thickness, and (iii) linear temperature assuming fixed $G=3.0$ (see Fig.~\ref{fig:temp_uni}(b)).
The evolution in the absence of Marangoni effects is shown as well - here, the film destabilizes on the time scale expected from the LSA.   
For self-consistent temperature computations, $h_m$ shows oscillatory behavior, in contrast to
the other considered approaches for temperature computations, or when Marangoni effect is 
not included.  Note that the numerical solution uses the substrate thickness $h_s = 100$; for such $h_s$, there is an
excellent agreement between the numerical and analytical temperature solutions for fixed film thickness, see Fig.~\ref{fig:temp_uni}(a).
Therefore, {\it the difference between the solutions is not due to analytical solution not being accurate, but due to the fact
that it ignores evolution of the film itself. }

Our finding so far is that Marangoni effect, when included self-consistently into Eq.~(\ref{eq:evol}), changes dramatically
the behavior of the film, leading to stabilization for the present choice of parameters.
The effect is particularly strong for thin films, that are strongly unstable due
to destabilizing disjoining pressure, if Marangoni effect is excluded.   The obvious question 
is whether these results are general, in particular in the light of experimental findings that find 
instability, see, e.g.,~\cite{trice_prl08,lang13}.     To start answering this question, we consider
the influence of two parameters: time dependence of the source term, and its total energy.
The influence of the domain size and of the number of physical dimensions is discussed later in Sec.~\ref{sec:large}.  
Further more detailed study of the influence of other parameters will be given elsewhere~\cite{in_prep}. 

\begin{figure}[t!]
\includegraphics[width= 0.48 \textwidth]{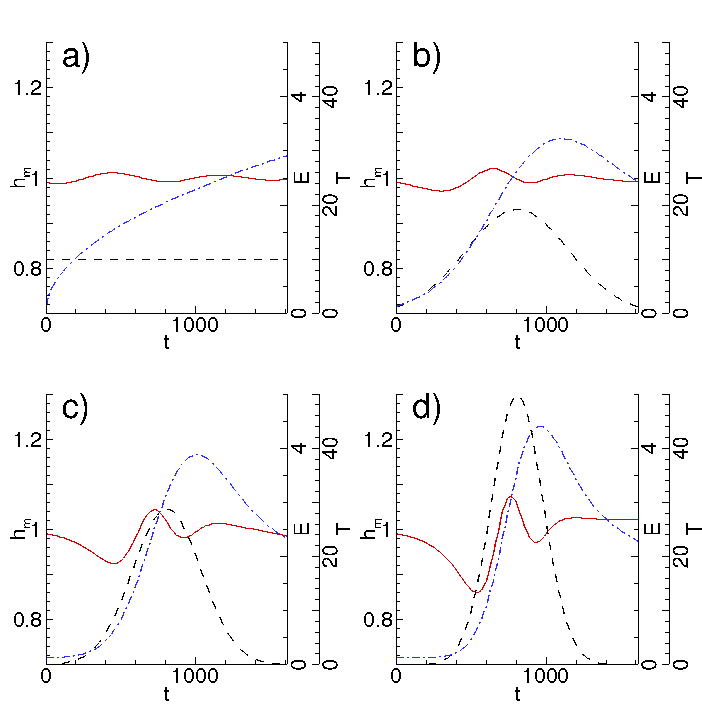}
\caption{\label{fig:sigma_high}
The film thickness, $h_m$ (red solid), 
numerically computed film temperature at $x_m$ (blue dash dotted), and 
the applied energy distribution (black dashed).   The total energy density applied, $E_0$,
during considered time window  is kept constant.
}
\end{figure}

Figure~\ref{fig:sigma_high} shows $h_m$ obtained by assuming Gaussian profile (function $F(t)$) of the source of 
different widths (see Appendix~\ref{sec:heat} for more details), keeping the total energy density the same as for the uniform profile considered so far.  
From Fig.~\ref{fig:sigma_high}, we 
observe that as the energy distribution of the source becomes more narrow, the oscillatory behavior of $h_m$ becomes stronger;
however we always find that the final outcome is consistent with the one obtained for a uniform source - stable film.   

\begin{figure}[t!]
\includegraphics[width= 0.52 \textwidth]{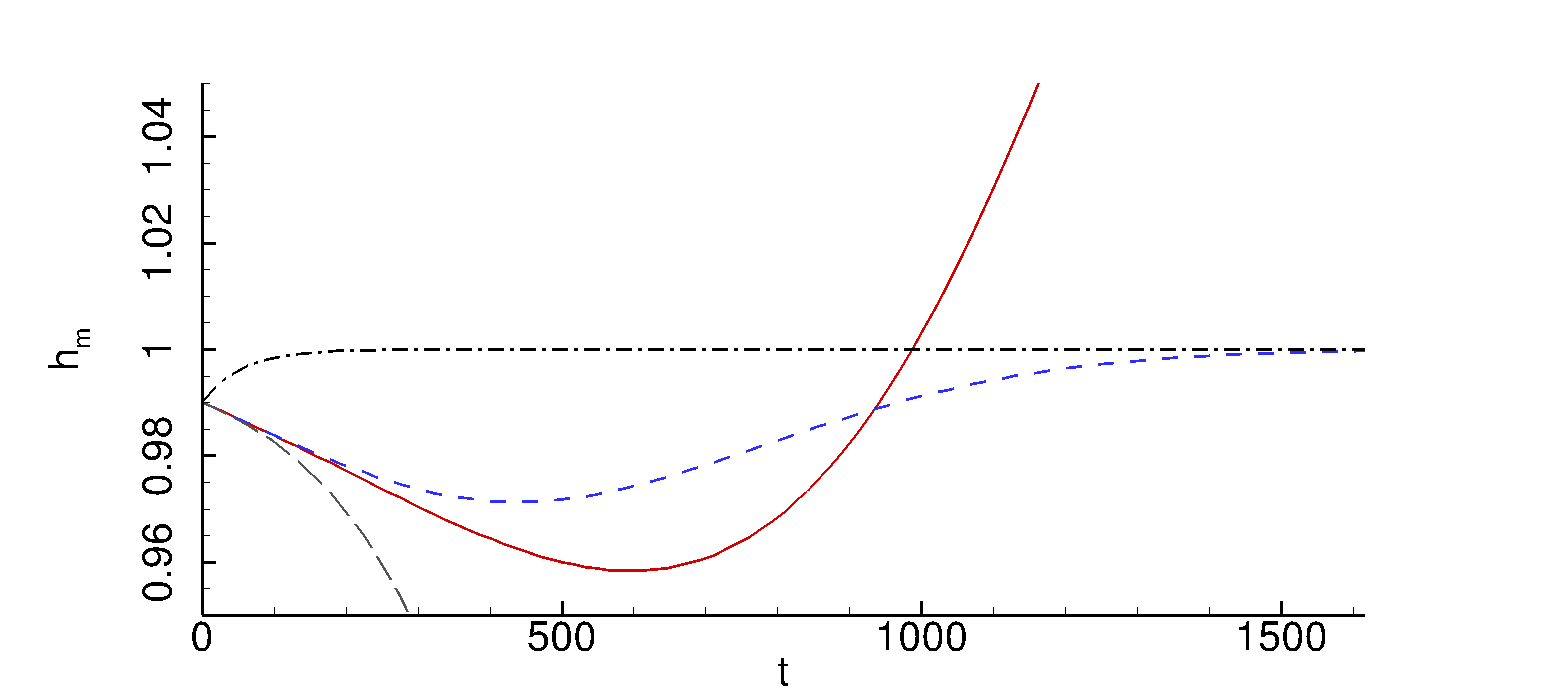}
\caption{\label{fig:h_evol_uni_l}
The film thickness, $h_m$, at $x_m = \lambda_m/2$ using different approaches to compute the temperature:  
self-consistent time-dependent temperature computation (red solid), the analytical solution of
Eq.~(\ref{eq:heat}) assuming fixed film thickness (blue dashed), and using fixed $G = 0.9$ obtained
from Fig.~\ref{fig:temp_uni}c) (black dash dotted).
The evolution computed by ignoring Marangoni effect all together is shown as well (grey long dashed).
Compare with Fig.~\ref{fig:h_evol_uni} where applied energy density, and the corresponding value of $G$, are larger.
}
\end{figure}

\begin{figure}[t!]
\includegraphics[width= 0.48 \textwidth]{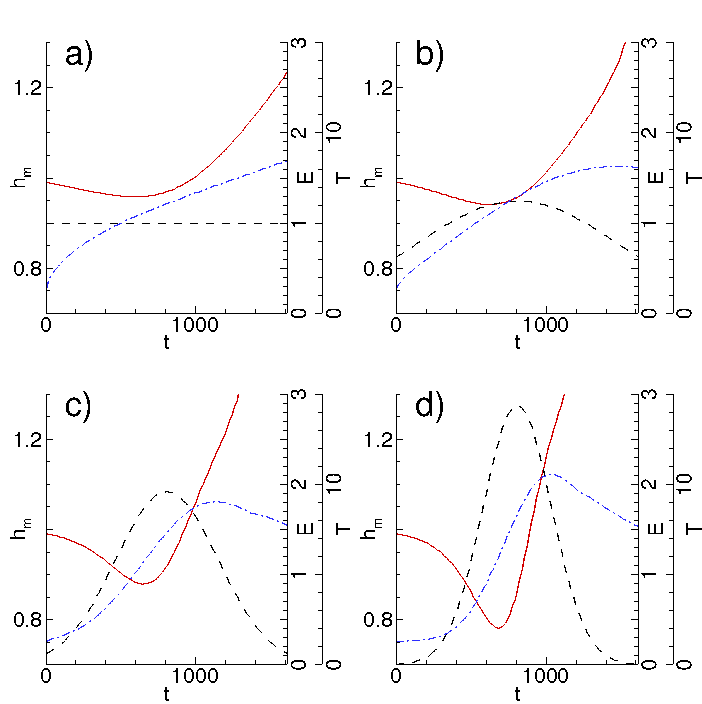}
\caption{\label{fig:sigma_low}
The film thickness, $h_m$ (red solid), 
numerically computed film temperature at $x_m$ (blue dash dotted), and 
the applied energy distribution (black dashed).   The total energy density applied
is kept constant at $E_0/4$ (one quarter of the one used in Fig.~\ref{fig:sigma_high}).
Note different scales for $T$ and $E$ compared to Fig.~\ref{fig:sigma_high}.}
\end{figure}

Next we consider the influence of the energy density of the source term on the evolution, keeping all other parameters
the same.  Figures~\ref{fig:h_evol_uni_l} and~\ref{fig:sigma_low} show the results obtained for the same setup as the one used for
Figs.~\ref{fig:h_evol_uni} and~\ref{fig:sigma_high} but with 
decreased energy density of the pulse.  Now, the evolution is unstable: while Marangoni effect is strong enough to suppress
initial instability growth (decrease of $h_m$), it is insufficient to stabilize the rebound: $h_m$ increases monotonously for later
times, with the  final outcome (for longer times than shown in Figs.~\ref{fig:h_evol_uni_l} and~\ref{fig:sigma_low}) of the 
formation of a drop centered at $x_m$.   Other outcomes are possible: e..g, for the total energy at some intermediate level   
between the ones used in Figs.~\ref{fig:sigma_high} and~\ref{fig:sigma_low}, one can find drops centered at the 
domain boundaries (results not shown for brevity).  

To summarize, we find that Marangoni effect can have profound effect on stability of a thin film on thermally conductive substrate, 
and may result in oscillatory decay or growth of free surface instability. We have focused here on the influence that the source term properties
have on the results; the influence of other ingredients in the model will be explored elsewhere~\cite{in_prep}.  In what follows, we 
discuss the influence of the domain size and its dimensionality.

\subsection{Marangoni effect for evolving films in large domains in two and three 
spatial dimensions}
\label{sec:large}

So far we have considered influence of Marangoni effect in small computational 
domains and in two spatial dimensions.  One may wonder whether freely evolving
films, unconstrained by domain size would evolve differently, particularly in 3D.   In
this section we consider evolution in large domains in both 2D and 3D and show 
that all the main conclusions that we have already reached  remain valid; in particular, the stability 
properties of the films remain as we have  already discussed.     For brevity, we will discuss only
the results obtained by fully self-consistent computation of the temperature field and resulting 
Marangoni effect, in addition to presenting the results of simulations that exclude Marangoni 
effects.   Since we have not so far found strong effect of the time dependence of the source 
term on the evolution, we will consider only uniform source here; however, since we 
observed that the total applied energy density does influence the result, we will include the 
results for the two energy densities considered so far in this section as well.

The results that follow focus on the same time range and source properties as considered so far; this 
is necessary so to avoid confusion regarding total energy density provided by the 
heat source in Eq.~(\ref{eq:heat}).  For this reason, some of the figures in 
this section (as in the preceding ones) present films that are still evolving.   In the
context of metal films, where films solidify after a laser pulse, any of the shown 
configurations may be a final one.  Longer time evolution that leads to formation 
of drops for all considered unstable configurations, as well as the evolution for multiple
laser pulses that are commonly used in experiments~\cite{trice_prl08,lang13,NanoLett14} will be considered in 
future work~\cite{in_prep}.

We consider the domain sizes that are equal to $20 \lambda_m$ (with $\lambda_m$ given 
by Eq.~(\ref{eq:LSA}) with Marangoni effect excluded) in
2D, or to $[5\lambda_m, 5\lambda_m]$ in 3D.   The 3D simulations are carried out by implementing the ADI method 
that has been already used in similar contexts, see e.g.,~\cite{lin_pof12,lin_jfm13,gdk_jfm13} for
examples, as well as~\cite{witelski_anm03} for a 
careful discussion of this method in the context of 4th order nonlinear diffusion equations.   
The boundary conditions are analogous to the
2D case, with the first and third derivatives vanishing in the direction normal to the 
domain boundary.

 The initial condition consists of a film perturbed by a set of random perturbations, specified
 as follows.     Consider $ N\times N$ grid in the 2D plane, with $z_{l,m}$ a random complex number of 
 unit length.  The initial condition is then specified by
\begin{equation}
 h_{k,j} = h_0 + \epsilon  \left| \sum^N_{l = 0} \left(e^{i2\pi k l/N} \sum^N_{m = 0} e^{i2\pi j m/N} z_{l,m}\right) \right| \ .
 \label{eq:ic}
\end{equation}
Here, $N$ is the number of grid points in each direction, and $\epsilon$ is the amplitude of perturbation.  
We use $\epsilon = 0.01$ and in 2D use the 1D version of Eq.~(\ref{eq:ic}).    Note that after this
initial stochastic perturbation, the evolution
is fully deterministic.  See~\cite{nesic_pre15,diez_pre16} for further discussion of fully stochastic 
evolution in the context of thin film dynamics. 

\begin{figure}[t!]
\includegraphics[width= 0.52\textwidth]{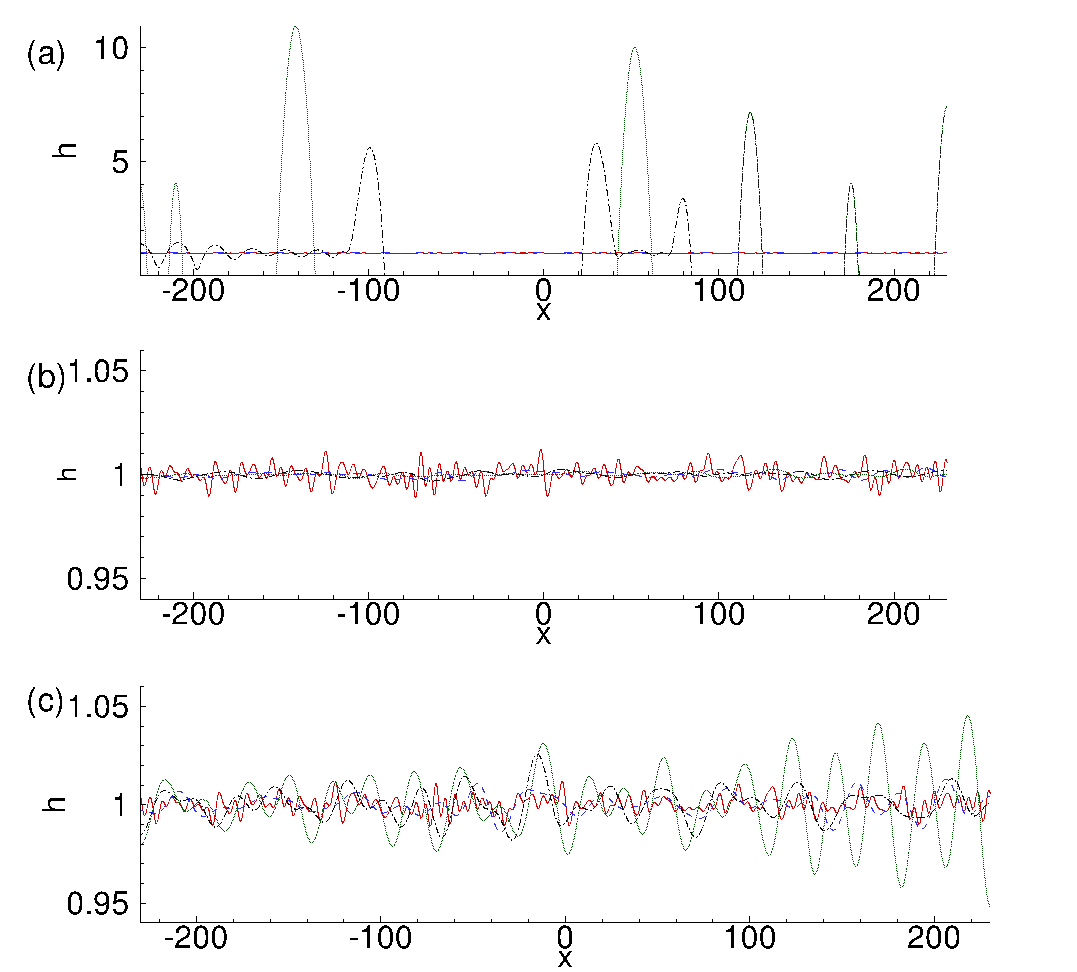}
\caption{
The evolution of a perturbed film in a large domain ($20$ times the size shown in Fig. \ref{fig:profiles_uni}),
the rest of the setup is the same as the one in Fig. \ref{fig:sigma_high} and Fig. \ref{fig:sigma_low}, with the difference that
in a) the film evolves without Marangoni effect; in b) the energy density of the source term is 
$E_0$, the same as in Fig. \ref{fig:sigma_high}; and in c) the energy density is $E_0/4$, the same
as in Fig. \ref{fig:sigma_low}. The initial condition is identical for all three figures and is shown by red solid lines, with the other 
lines showing the results at $t = 509$  (blue dashed), $t = 1074$ (black dashdot), and  $t = 1583$ (green dotted).
\label{fig:2D_large}
}
\end{figure}

Figure~\ref{fig:2D_large} shows the evolution of randomly perturbed 2D film for
three cases: without Marangoni effect (a), and with Marangoni effect included, 
and the total energy density applied equal to $E_0$ (b), and $E_0/4$ (c).   We observe
very different evolutions, with the instability growing quickly in (a), oscillatory 
instability decay in (b) and oscillatory instability growth in (c), consistently
with the LSA for the no-Marangoni case, and with the results obtained in small
computational domain and  a single perturbation shown in the preceeding figures. 

During the evolution time shown in Fig.~\ref{fig:2D_large}, the instability has only started to 
grow (part b)) or decay (part c)), but has already led to the formation of drops in the part a), for 
which Marangoni effect is excluded.   This finding  (implicit in the earlier figures as well) suggests
that instability with and without Marangoni effect evolves on different time scales, with 
faster evolution if Marangoni effect is not considered.   The question 
of the influence of Marangoni effect on emerging length scales in the nonlinear regime of 
instability development is not simple to answer and its careful consideration is referred to 
future work~\cite{in_prep}.    Already for evolution without Marangoni effect, 
shown in Fig.~\ref{fig:2D_large}, we note rather strong coarsening effect - the distance between
the drops that are about to form is much larger than the most unstable wavelength, $\lambda_m$, 
obtained from the LSA.     This coarsening is consistent with the results of simulations focusing 
on stochastic effects~\cite{nesic_pre15} and has not been, to our knowledge, carefully analyzed
yet.  

\begin{figure}[t!]
\includegraphics[width= 0.52\textwidth]{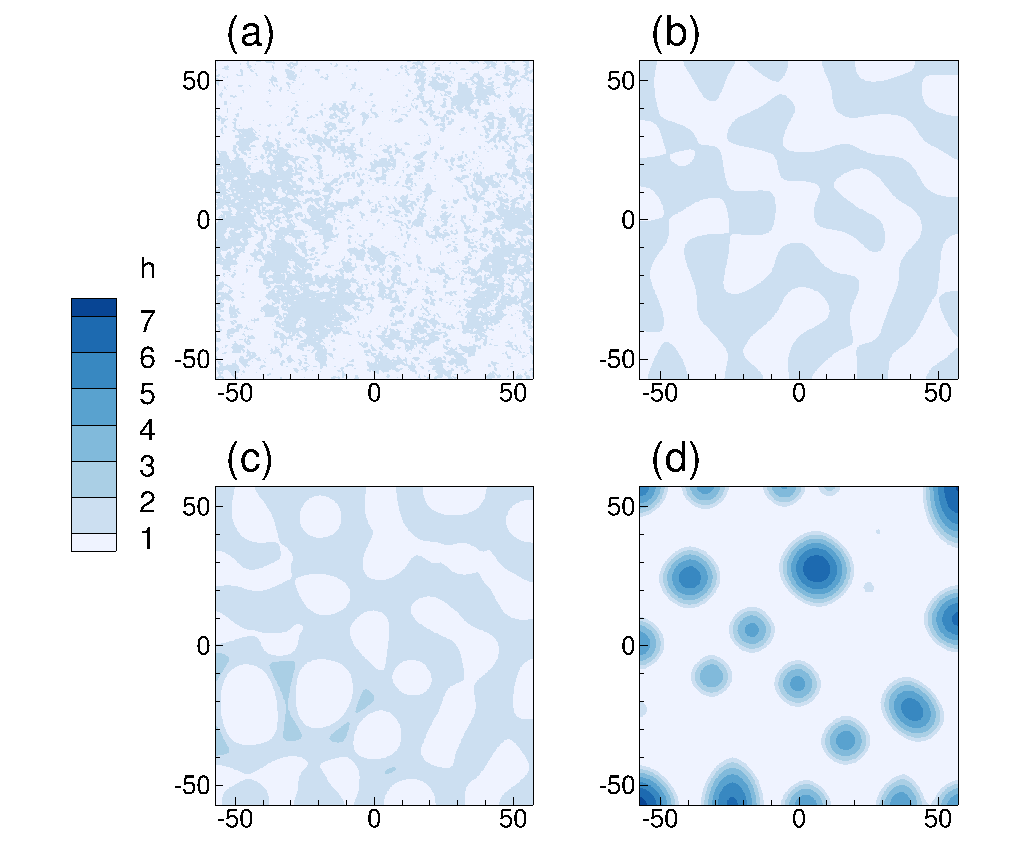}
\caption{\label{fig:3DnoMaran}
The evolution of a flat film with random initial perturbation in 3D geometry.  
Marangoni effect is not considered.    The results in this figure as well as in Figs.~\ref{fig:3D_large_e} and~\ref{fig:3D_large_e4}
are shown at  (a) $t = 0$; (b) $t = 528$, (c) $t = 1040$, and (d) $t = 1578$.   
\label{fig:3D_large_nm}
}
\end{figure}

\begin{figure}[t!]
\includegraphics[width= 0.52\textwidth]{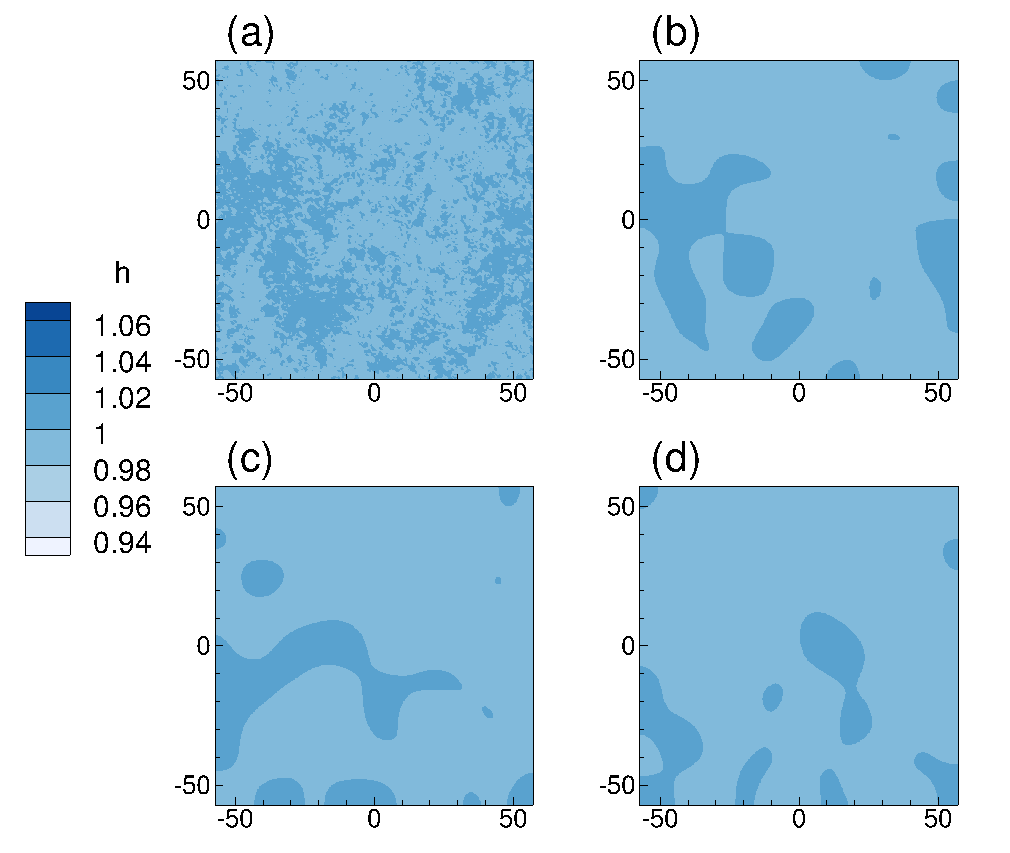}
\caption{\label{fig:3DE468}
The evolution of a flat film with random initial perturbation in 3D geometry. 
Marangoni effect is included, with the total energy density applied corresponding 
to $E_0$.    
\label{fig:3D_large_e}
}
\end{figure}

\begin{figure}[t!]
\includegraphics[width= 0.52\textwidth]{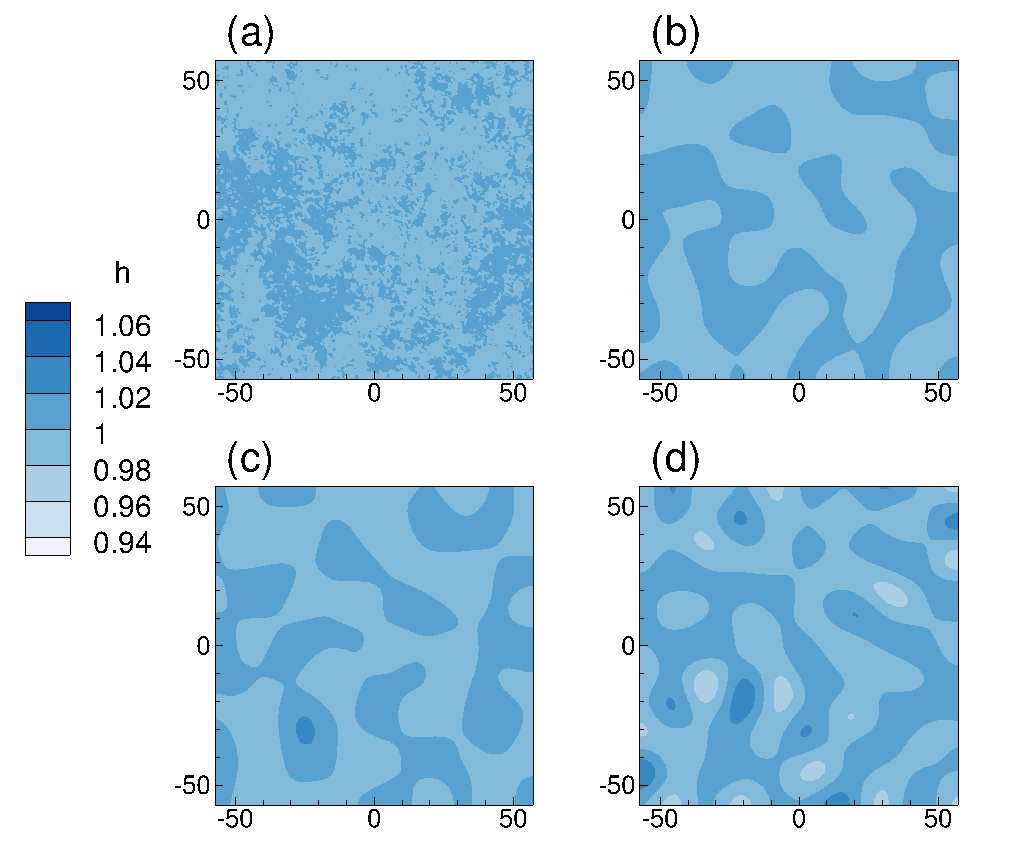}
\caption{\label{fig:3DE1875}
The evolution of a flat film with random initial perturbation in 3D geometry. 
Marangoni effect is included, with the total energy density applied corresponding 
to $E_0/4$.   
\label{fig:3D_large_e4}
}
\end{figure}

Next we proceed with 3D simulations.   
Figure~\ref{fig:3D_large_nm} shows the results obtained in simulations that do not 
include Marangoni effect, corresponding to the ones shown in Fig.~\ref{fig:2D_large}(a); 
Fig.~\ref{fig:3D_large_e} and~\ref{fig:3D_large_e4} then show the results for 
the total energy density equal to $E_0$ and $E_0/4$, corresponding to 
Fig.~\ref{fig:2D_large}(b) and (c), respectively.   Figure~\ref{fig:3D_large_nm}, where 
final drops have already formed, shows similar coarsening effect as in 2D; the typical 
length scale (distance between the drops) is  larger that $\lambda_m$ obtained 
based on the LSA; similar coarsening was found when analyzing experimental 
results for unstable Cu films~\cite{lang13}.  

When Marangoni effect is considered, as in Figs.~\ref{fig:3D_large_e}~-~\ref{fig:3D_large_e4}, 
the evolution is consistent with the 2D versions shown in Fig.~\ref{fig:2D_large}(b)~-~(c).  In 
particular, the stability properties of the films are not influenced by the geometry: the films 
that are unstable in 2D are unstable in 3D as well.   These  results show clearly strong influence 
of Marangoni effect on the instability evolution, suggesting that temperature dependence of 
surface tension may be used to control the instability development in physical experiments.   

In the present work, we focus on the early stages of instability development, particularlly in the 
case of evolution  in the presence of Marangoni effect.  Our future work~\cite{in_prep} will discuss 
in more detail the evolution at the later stages that involve formation of drops, and the influence
of Marangoni effect on the emerging length scales.

\section{Conclusions}
\label{sec:conclusions}

The main conclusion of this work is that careful consideration of heat conductions is required to properly
account for the influence of Marangoni forces on the film evolution: in particular, the assumption that the film 
temperature is slaved to its thickness leads to different results from the ones obtained by self-consistent 
computations.    The sensitivity of the outcome as the parameters entering the problem (such as total energy density of
the source term)  are modified suggests that more general insight 
could be reached by carrying out further studies using more elaborate linear and weakly-nonlinear 
analyses of the evolution.  We hope that our results will inspire further research in this direction.

We note that in the present work we have considered a very basic model, and have not included 
a number of effects: solidification and melting are not considered, modeling of heat flow is limited
to one dimension; the substrate itself is considered uniform, and the other physical parameters (such 
as viscosity and thermal conductivity) are considered to be temperature-independent.  We expect that
the results presented here will serve as a basis for further improvements, in particular since they show 
that Marangoni effect may influence strongly both time-scales and length-scales of instability development,
opening the door to its use for the purpose of controlled directed assembly on nanoscale.  Our research will
proceed in this direction.

\section*{Acknowledgements} 
The authors thank Shahriar Afkhami, Jason Fowlkes, Kyle Mahady, Philip Rack, and  Ivana Seric for many insightful discussions.
This work was partially supported by the NSF Grant No. CBET-1235710.

\appendix
\section{Parameters}\label{sec:param}

Table \ref{table:ref_paras} provides the parameters and scales used in the main text. The film parameters (subscript `m')
assume Cu film, and the substrate parameters (subscript `s') assume SiO$_2$.   

\begin{center} 
\begin{table}[h]
  \begin{tabular}{ | l | l | l |} \hline 
  Parameter & Value & Unit  \\ \hline 
  viscosity ($\mu$) & $4.3\times 10^{-3}$ & $\mathrm{m^2/s}$  \\ \hline 
  surface tension ($\gamma$) & 1.303 & $\mathrm{J/m^2}$ \\ \hline 
  length scale ($l_{s}$) & $1.0\times 10^{-8}$ & $\mathrm{m}$  \\ \hline 
  time scale ($t_{s} = 3l_{s}\mu/\gamma$) & $9.21\times 10^{-11}$ & $\mathrm{s}$ \\ \hline 
  film density ($\rho_{m}$) & $8.0\times 10^{3}$ & $\mathrm{kg/m^3}$  \\ \hline 
  SiO$_2$ density ($\rho_\mathrm{SiO2}$)  & $2.2\times 10^{3}$ & $\mathrm{kg/m^3}$  \\ \hline
  film heat capacity ($\mathrm{Ceff}_\mathrm{m}$) & $4.95\times 10^{2}$ & $\mathrm{J/kg/K}$ \\ \hline 
  SiO$_2$ heat capacity ($\mathrm{Ceff}_\mathrm{SiO2}$) & $9.37\times 10^{2}$ & $\mathrm{J/kg/K}$ \\ \hline 
  film heat conductivity ($k_{m}$) & $3.40\times 10^{2}$ & $\mathrm{W/m/K}$  \\ \hline
  film absorption length ($\alpha_m^{-1}$) & $11.09\times 10^{-9}$ & $\mathrm{m}$  \\ \hline
  SiO$_2$ heat conductivity ($k_{SiO2}$) & $1.4\times 10^{0}$ & $\mathrm{W/m/K}$\\ \hline 
  surface tension dep of T ($\gamma_T$) & $-2.3\times 10^{-4}$ & $\mathrm{J/m^2}$  \\ \hline 
    Avogadro's constant ($A$) & $1.83\times 10^{-18}$ & $\mathrm{J}$  \\ \hline   
    reflective coefficient ($r_0$) & $0.3655$ & 1  \\ \hline
  film reflective length ($\alpha_r^{-1}$) & $12.0\times 10^{-9}$ & $\mathrm{m}$ \\ \hline 
    laser energy density ($E_0$) & $8.80\times 10^{3}$ & $\mathrm{J/m^2}$  \\ \hline
    time duration of observation ($t_\mathrm{total}$) & 160 & ns  \\ \hline
    Gaussian pulse peak time ($t_p$) & 80 & ns \\ \hline
    equilibrium film thickness ($h_*$) & $1.0\times 10^{-10}$ & $\mathrm{m}$  \\ \hline 
  film thickness ($h_0$) & $1.0\times 10^{-8}$ & m \\ \hline 
  SiO$_2$ thickness ($h_\mathrm{SiO2}$) & $1.0\times 10^{-6}$ & m  \\ \hline
  room temperature ($T_\mathrm{room}$) & $300$ & K \\ \hline 
 \end{tabular} 
 \caption{The parameters used in the main text.}
 \label{table:ref_paras}
\end{table}
\end{center}

\section{Formulation of the heat diffusion problem}
\label{sec:heat}

The dimensional heat diffusion equation with the time dependent source term, considered in the main text, is as follows
\begin{equation}\label{eq:1D-heat_m}
 (\rho C_{eff})_m \frac{\partial T}{\partial t} = k_m \frac{\partial^2 T}{\partial z^2} + S^* F(t) \alpha_m e^{-\alpha_m(z-h)} \ .
\end{equation}
Here we take the film  surface to be at $z=h$, and the film-substrate interface is at $z=0$.
In the substrate layer, the heat absorption is ignored, leading to 
\begin{equation}\label{eq:1D-heat_s}
 (\rho C_{eff})_s \frac{\partial T}{\partial t} = k_s \frac{\partial^2 T}{\partial z^2} \ .
\end{equation}
The scales and parameters are given in Table~\ref{table:ref_paras}.
For 
uniform pulse we have
$$
 S^* = [1-R(h)]\frac{E_0}{t_p}; \quad F(t) = 1\ ,
$$
where $R(h)$ is the overall material reflectivity. 
$$
 R(h) = r_0 (1-e^{-\alpha_r h})\ ,
$$
For Gaussian pulse, 
$$
 S^* = [1-R(h)]\frac{E_0 \zeta}{\sqrt{2\pi}\sigma}; \quad F(t) = \exp(-(t-t_p)^2/\sigma^2)\ ,
$$
Here, $\zeta$ is a renormalization factor used to ensure that during the considered observation time, $t_\mathrm{total}$,
the Gaussian and the uniform pulse lead to the same total applied energy.

Equation (2) from the main body of the text is obtained by using
the length and time scale as defined there, and the temperature 
scale $T_s = {t_s E_0 \alpha_m/(\rho C_\mathrm{eff})_m t_p)}$:
$$
 K_1 = \frac{k_m t_s}{(\rho C_\mathrm{eff})_\mathrm{m} l_s^2}   \ ,
$$

$$
 Q_1 = \frac{t_s S^* F(t) \alpha_m e^{-\alpha_m(z-h)}}{(\rho C_{eff})_m T_s}\ , 
$$

$$
  K_2 = \frac{k_s t_s}{(\rho C_\mathrm{eff})_\mathrm{s} l_s^2}; \quad Q_2 = 0\ .
$$

\section{Outline of the derivation of the analytical solution to the heat diffusion problem }
\label{sec:analytical}

Here, we give a brief overview of the derivation of the analytical solution of 
Eqs. (\ref{eq:1D-heat_m}-\ref{eq:1D-heat_s}), assuming that the film thickness 
is constant (or, equivalently, that the temperature is slaved to the current value of the 
film thickness), assuming also that the substrate layer 
is infinitely thick.   This formulation was discussed in more details in~\cite{trice_prb07}, 
and also used for the purpose of estimating liquid lifetime of metal film in~\cite{fowlkes_nano11,NanoLett14,langmuir15}.   
Let 
$$
 S = \frac{S^*[1-e^{-\alpha_m h}]}{(\rho C_{eff}) h}; \quad K = \frac{\sqrt{(\rho C_\mathrm{eff}k)_s}}{(\rho C_\mathrm{eff})_m)h}\ ,
$$
and 
$$
 q_s(t) = -k_m (\partial T/\partial z)_m = -k_s(\partial T/\partial z)_s\ .
$$
Here $q_s(t)$ represents the heat flux through the film-substrate interface.  Since the film layer is thin and the
heat conduction high, the time scale 
for heat conduction in the $z$ direction is short, $\approx 10^{-2}$ns using the parameters as given in Table~\ref{table:ref_paras}. 
There, the approximation that $T\neq f(z)$ is expected to be highly accurate.
Integrating Eq. (\ref{eq:1D-heat_m}) from $z=h$ to $z=0$ and taking the average, we find
\begin{equation}
 T(t) = T_0+\int_0^t \left(S f(\tau)-\frac{q_s(\tau)}{(\rho C_{eff})_m h}\right) d\tau \ .
 \label{eq:temp}
\end{equation}
Solving the heat equation in the substrate of semi-infinite thickness gives
$$ T_s(t,z) = T_0 + \frac{\sqrt{\alpha_s}}{k_s \sqrt{\pi}} \int_0^t q_s(t)(t-\tau)^{1\over 2} \exp\left(\frac{-z^2}{4\alpha_s(t-\tau)}\right)d\tau \ .
$$
Using $T(t)=T_s(t,0)$ we have
$$
 S\int_0^t f(\tau)d\tau = \int_0^t \frac{q_s(\tau)d\tau}{(\rho C_\mathrm{eff})_m h}
 +\frac{\sqrt{\alpha_s}}{k_s\sqrt{\pi}}\int_0^t \frac{q_s(\tau)}{\sqrt{t-\tau}}d\tau\ .
$$
Using Laplace transform, we can solve for $q_s(t)$, and substituting the result into Eq.~(\ref{eq:temp}), we 
obtain the solution for the film temperature
\begin{equation}
 T(t) = T_0+S \int_0^t e^{K^2 u} \mathrm{erfc}(K\sqrt{u})d u\ .
\end{equation}
This $T(t)$ is shown in the main body of the paper as the analytical solution.  
Note that its derivation assumes that $h$ remains constant in time.

\bibliography{films}

\begin{thebibliography}{30}%
\makeatletter
\providecommand \@ifxundefined [1]{%
 \@ifx{#1\undefined}
}%
\providecommand \@ifnum [1]{%
 \ifnum #1\expandafter \@firstoftwo
 \else \expandafter \@secondoftwo
 \fi
}%
\providecommand \@ifx [1]{%
 \ifx #1\expandafter \@firstoftwo
 \else \expandafter \@secondoftwo
 \fi
}%
\providecommand \natexlab [1]{#1}%
\providecommand \enquote  [1]{``#1''}%
\providecommand \bibnamefont  [1]{#1}%
\providecommand \bibfnamefont [1]{#1}%
\providecommand \citenamefont [1]{#1}%
\providecommand \href@noop [0]{\@secondoftwo}%
\providecommand \href [0]{\begingroup \@sanitize@url \@href}%
\providecommand \@href[1]{\@@startlink{#1}\@@href}%
\providecommand \@@href[1]{\endgroup#1\@@endlink}%
\providecommand \@sanitize@url [0]{\catcode `\\12\catcode `\$12\catcode
  `\&12\catcode `\#12\catcode `\^12\catcode `\_12\catcode `\%12\relax}%
\providecommand \@@startlink[1]{}%
\providecommand \@@endlink[0]{}%
\providecommand \url  [0]{\begingroup\@sanitize@url \@url }%
\providecommand \@url [1]{\endgroup\@href {#1}{\urlprefix }}%
\providecommand \urlprefix  [0]{URL }%
\providecommand \Eprint [0]{\href }%
\providecommand \doibase [0]{http://dx.doi.org/}%
\providecommand \selectlanguage [0]{\@gobble}%
\providecommand \bibinfo  [0]{\@secondoftwo}%
\providecommand \bibfield  [0]{\@secondoftwo}%
\providecommand \translation [1]{[#1]}%
\providecommand \BibitemOpen [0]{}%
\providecommand \bibitemStop [0]{}%
\providecommand \bibitemNoStop [0]{.\EOS\space}%
\providecommand \EOS [0]{\spacefactor3000\relax}%
\providecommand \BibitemShut  [1]{\csname bibitem#1\endcsname}%
\let\auto@bib@innerbib\@empty
%</preamble>
\bibitem [{\citenamefont {Oron}\ \emph {et~al.}(1997)\citenamefont {Oron},
  \citenamefont {Davis},\ and\ \citenamefont {Bankoff}}]{oron_rmp97}%
  \BibitemOpen
  \bibfield  {author} {\bibinfo {author} {\bibfnamefont {A.}~\bibnamefont
  {Oron}}, \bibinfo {author} {\bibfnamefont {S.~H.}\ \bibnamefont {Davis}}, \
  and\ \bibinfo {author} {\bibfnamefont {S.~G.}\ \bibnamefont {Bankoff}},\
  }\href@noop {} {\bibfield  {journal} {\bibinfo  {journal} {Rev. Mod. Phys.}\
  }\textbf {\bibinfo {volume} {69}},\ \bibinfo {pages} {931} (\bibinfo {year}
  {1997})}\BibitemShut {NoStop}%
\bibitem [{\citenamefont {Craster}\ and\ \citenamefont
  {Matar}(2009)}]{cm_rmp09}%
  \BibitemOpen
  \bibfield  {author} {\bibinfo {author} {\bibfnamefont {R.}~\bibnamefont
  {Craster}}\ and\ \bibinfo {author} {\bibfnamefont {O.}~\bibnamefont
  {Matar}},\ }\href@noop {} {\bibfield  {journal} {\bibinfo  {journal} {Rev.
  Mod. Phys.}\ }\textbf {\bibinfo {volume} {81}},\ \bibinfo {pages} {1131}
  (\bibinfo {year} {2009})}\BibitemShut {NoStop}%
\bibitem [{\citenamefont {Colinet}\ \emph {et~al.}(2001)\citenamefont
  {Colinet}, \citenamefont {Legros},\ and\ \citenamefont {Velarde}}]{COLINET}%
  \BibitemOpen
  \bibfield  {author} {\bibinfo {author} {\bibfnamefont {P.}~\bibnamefont
  {Colinet}}, \bibinfo {author} {\bibfnamefont {J.}~\bibnamefont {Legros}}, \
  and\ \bibinfo {author} {\bibfnamefont {M.}~\bibnamefont {Velarde}},\
  }\href@noop {} {\emph {\bibinfo {title} {{Nonlinear dynamics of
  surface-tension-driven instabilities}}}}\ (\bibinfo  {publisher}
  {Wiley-VCH},\ \bibinfo {address} {Berlin},\ \bibinfo {year}
  {2001})\BibitemShut {NoStop}%
\bibitem [{\citenamefont {Podolny}\ \emph {et~al.}(2005)\citenamefont
  {Podolny}, \citenamefont {Oron},\ and\ \citenamefont
  {Nepomnyashchy}}]{podolny05}%
  \BibitemOpen
  \bibfield  {author} {\bibinfo {author} {\bibfnamefont {A.}~\bibnamefont
  {Podolny}}, \bibinfo {author} {\bibfnamefont {A.}~\bibnamefont {Oron}}, \
  and\ \bibinfo {author} {\bibfnamefont {A.~A.}\ \bibnamefont
  {Nepomnyashchy}},\ }\href@noop {} {\bibfield  {journal} {\bibinfo  {journal}
  {Phys. Fluids}\ }\textbf {\bibinfo {volume} {17}},\ \bibinfo {eid} {104104}
  (\bibinfo {year} {2005})}\BibitemShut {NoStop}%
\bibitem [{\citenamefont {Morozov}\ \emph {et~al.}(2015)\citenamefont
  {Morozov}, \citenamefont {Oron},\ and\ \citenamefont
  {Nepomnyashchy}}]{morozov15}%
  \BibitemOpen
  \bibfield  {author} {\bibinfo {author} {\bibfnamefont {M.}~\bibnamefont
  {Morozov}}, \bibinfo {author} {\bibfnamefont {A.}~\bibnamefont {Oron}}, \
  and\ \bibinfo {author} {\bibfnamefont {A.~A.}\ \bibnamefont
  {Nepomnyashchy}},\ }\href@noop {} {\bibfield  {journal} {\bibinfo  {journal}
  {Phys. Fluids}\ }\textbf {\bibinfo {volume} {27}},\ \bibinfo {eid} {082107}
  (\bibinfo {year} {2015})}\BibitemShut {NoStop}%
\bibitem [{\citenamefont {Nepomnyashchy}\ and\ \citenamefont
  {Simanovskii}(2015)}]{nepom_jfm15}%
  \BibitemOpen
  \bibfield  {author} {\bibinfo {author} {\bibfnamefont {A.}~\bibnamefont
  {Nepomnyashchy}}\ and\ \bibinfo {author} {\bibfnamefont {I.}~\bibnamefont
  {Simanovskii}},\ }\href {\doibase 10.1017/jfm.2015.178} {\bibfield  {journal}
  {\bibinfo  {journal} {J. Fluid Mech.}\ }\textbf {\bibinfo {volume} {771}},\
  \bibinfo {pages} {159} (\bibinfo {year} {2015})}\BibitemShut {NoStop}%
\bibitem [{\citenamefont {Shklyaev}\ \emph {et~al.}(2010)\citenamefont
  {Shklyaev}, \citenamefont {Khenner},\ and\ \citenamefont
  {Alabuzhev}}]{shklyaev10}%
  \BibitemOpen
  \bibfield  {author} {\bibinfo {author} {\bibfnamefont {S.}~\bibnamefont
  {Shklyaev}}, \bibinfo {author} {\bibfnamefont {M.}~\bibnamefont {Khenner}}, \
  and\ \bibinfo {author} {\bibfnamefont {A.~A.}\ \bibnamefont {Alabuzhev}},\
  }\href {\doibase 10.1103/PhysRevE.82.025302} {\bibfield  {journal} {\bibinfo
  {journal} {Phys. Rev. E}\ }\textbf {\bibinfo {volume} {82}},\ \bibinfo
  {pages} {025302} (\bibinfo {year} {2010})}\BibitemShut {NoStop}%
\bibitem [{\citenamefont {Shklyaev}\ \emph {et~al.}(2012)\citenamefont
  {Shklyaev}, \citenamefont {Alabuzhev},\ and\ \citenamefont
  {Khenner}}]{shklyaev12}%
  \BibitemOpen
  \bibfield  {author} {\bibinfo {author} {\bibfnamefont {S.}~\bibnamefont
  {Shklyaev}}, \bibinfo {author} {\bibfnamefont {A.~A.}\ \bibnamefont
  {Alabuzhev}}, \ and\ \bibinfo {author} {\bibfnamefont {M.}~\bibnamefont
  {Khenner}},\ }\href {\doibase 10.1103/PhysRevE.85.016328} {\bibfield
  {journal} {\bibinfo  {journal} {Phys. Rev. E}\ }\textbf {\bibinfo {volume}
  {85}},\ \bibinfo {pages} {016328} (\bibinfo {year} {2012})}\BibitemShut
  {NoStop}%
\bibitem [{\citenamefont {Samoilova}\ and\ \citenamefont
  {Lobov}(2014)}]{samoilova14}%
  \BibitemOpen
  \bibfield  {author} {\bibinfo {author} {\bibfnamefont {A.~E.}\ \bibnamefont
  {Samoilova}}\ and\ \bibinfo {author} {\bibfnamefont {N.~I.}\ \bibnamefont
  {Lobov}},\ }\href@noop {} {\bibfield  {journal} {\bibinfo  {journal} {Phys.
  Fluids}\ }\textbf {\bibinfo {volume} {26}},\ \bibinfo {eid} {064101}
  (\bibinfo {year} {2014})}\BibitemShut {NoStop}%
\bibitem [{\citenamefont {Warner}\ \emph {et~al.}(2002)\citenamefont {Warner},
  \citenamefont {Craster},\ and\ \citenamefont {Matar}}]{warner02}%
  \BibitemOpen
  \bibfield  {author} {\bibinfo {author} {\bibfnamefont {M.~R.~E.}\
  \bibnamefont {Warner}}, \bibinfo {author} {\bibfnamefont {R.~V.}\
  \bibnamefont {Craster}}, \ and\ \bibinfo {author} {\bibfnamefont {O.~K.}\
  \bibnamefont {Matar}},\ }\href@noop {} {\bibfield  {journal} {\bibinfo
  {journal} {Phys. Fluids}\ }\textbf {\bibinfo {volume} {14}},\ \bibinfo
  {pages} {1642} (\bibinfo {year} {2002})}\BibitemShut {NoStop}%
\bibitem [{\citenamefont {Atena}\ and\ \citenamefont
  {Khenner}(2009)}]{atena09}%
  \BibitemOpen
  \bibfield  {author} {\bibinfo {author} {\bibfnamefont {A.}~\bibnamefont
  {Atena}}\ and\ \bibinfo {author} {\bibfnamefont {M.}~\bibnamefont
  {Khenner}},\ }\href {\doibase 10.1103/PhysRevB.80.075402} {\bibfield
  {journal} {\bibinfo  {journal} {Phys. Rev. B}\ }\textbf {\bibinfo {volume}
  {80}},\ \bibinfo {pages} {075402} (\bibinfo {year} {2009})}\BibitemShut
  {NoStop}%
\bibitem [{\citenamefont {Khenner}\ \emph {et~al.}(2011)\citenamefont
  {Khenner}, \citenamefont {Yadavali},\ and\ \citenamefont
  {Kalyanaraman}}]{khenner_pof11}%
  \BibitemOpen
  \bibfield  {author} {\bibinfo {author} {\bibfnamefont {M.}~\bibnamefont
  {Khenner}}, \bibinfo {author} {\bibfnamefont {S.}~\bibnamefont {Yadavali}}, \
  and\ \bibinfo {author} {\bibfnamefont {R.}~\bibnamefont {Kalyanaraman}},\
  }\href@noop {} {\bibfield  {journal} {\bibinfo  {journal} {Phys. Fluids.}\
  }\textbf {\bibinfo {volume} {23}},\ \bibinfo {pages} {122105} (\bibinfo
  {year} {2011})}\BibitemShut {NoStop}%
\bibitem [{\citenamefont {Trice}\ \emph {et~al.}(2008)\citenamefont {Trice},
  \citenamefont {Thomas}, \citenamefont {Favazza}, \citenamefont
  {Sureshkumar},\ and\ \citenamefont {Kalyanaraman}}]{trice_prl08}%
  \BibitemOpen
  \bibfield  {author} {\bibinfo {author} {\bibfnamefont {J.}~\bibnamefont
  {Trice}}, \bibinfo {author} {\bibfnamefont {D.}~\bibnamefont {Thomas}},
  \bibinfo {author} {\bibfnamefont {C.}~\bibnamefont {Favazza}}, \bibinfo
  {author} {\bibfnamefont {R.}~\bibnamefont {Sureshkumar}}, \ and\ \bibinfo
  {author} {\bibfnamefont {R.}~\bibnamefont {Kalyanaraman}},\ }\href@noop {}
  {\bibfield  {journal} {\bibinfo  {journal} {Phys. Rev. Lett.}\ }\textbf
  {\bibinfo {volume} {101}},\ \bibinfo {pages} {017802} (\bibinfo {year}
  {2008})}\BibitemShut {NoStop}%
\bibitem [{\citenamefont {Ajaev}\ and\ \citenamefont
  {Willis}(2003)}]{ajaev_pof03}%
  \BibitemOpen
  \bibfield  {author} {\bibinfo {author} {\bibfnamefont {V.}~\bibnamefont
  {Ajaev}}\ and\ \bibinfo {author} {\bibfnamefont {D.}~\bibnamefont {Willis}},\
  }\href@noop {} {\bibfield  {journal} {\bibinfo  {journal} {{Phys. Fluids}}\
  }\textbf {\bibinfo {volume} {15}},\ \bibinfo {pages} {3144} (\bibinfo {year}
  {2003})}\BibitemShut {NoStop}%
\bibitem [{\citenamefont {Trice}\ \emph {et~al.}(2007)\citenamefont {Trice},
  \citenamefont {Thomas}, \citenamefont {Favazza}, \citenamefont
  {Sureshkumar},\ and\ \citenamefont {Kalyanaraman}}]{trice_prb07}%
  \BibitemOpen
  \bibfield  {author} {\bibinfo {author} {\bibfnamefont {J.}~\bibnamefont
  {Trice}}, \bibinfo {author} {\bibfnamefont {D.}~\bibnamefont {Thomas}},
  \bibinfo {author} {\bibfnamefont {C.}~\bibnamefont {Favazza}}, \bibinfo
  {author} {\bibfnamefont {R.}~\bibnamefont {Sureshkumar}}, \ and\ \bibinfo
  {author} {\bibfnamefont {R.}~\bibnamefont {Kalyanaraman}},\ }\href@noop {}
  {\bibfield  {journal} {\bibinfo  {journal} {Phys. Rev. B}\ }\textbf {\bibinfo
  {volume} {75}},\ \bibinfo {pages} {235439} (\bibinfo {year}
  {2007})}\BibitemShut {NoStop}%
\bibitem [{\citenamefont {Hartnett}\ \emph {et~al.}(2015)\citenamefont
  {Hartnett}, \citenamefont {Mahady}, \citenamefont {Fowlkes}, \citenamefont
  {Afkhami}, \citenamefont {Kondic},\ and\ \citenamefont {Rack}}]{langmuir15}%
  \BibitemOpen
  \bibfield  {author} {\bibinfo {author} {\bibfnamefont {C.~A.}\ \bibnamefont
  {Hartnett}}, \bibinfo {author} {\bibfnamefont {K.}~\bibnamefont {Mahady}},
  \bibinfo {author} {\bibfnamefont {J.~D.}\ \bibnamefont {Fowlkes}}, \bibinfo
  {author} {\bibfnamefont {S.}~\bibnamefont {Afkhami}}, \bibinfo {author}
  {\bibfnamefont {L.}~\bibnamefont {Kondic}}, \ and\ \bibinfo {author}
  {\bibfnamefont {P.~D.}\ \bibnamefont {Rack}},\ }\href@noop {} {\bibfield
  {journal} {\bibinfo  {journal} {Langmuir}\ }\textbf {\bibinfo {volume}
  {31}},\ \bibinfo {pages} {13609} (\bibinfo {year} {2015})}\BibitemShut
  {NoStop}%
\bibitem [{\citenamefont {Kondic}\ \emph {et~al.}(2009)\citenamefont {Kondic},
  \citenamefont {Diez}, \citenamefont {Rack}, \citenamefont {Guan},\ and\
  \citenamefont {Fowlkes}}]{kd_pre09}%
  \BibitemOpen
  \bibfield  {author} {\bibinfo {author} {\bibfnamefont {L.}~\bibnamefont
  {Kondic}}, \bibinfo {author} {\bibfnamefont {J.}~\bibnamefont {Diez}},
  \bibinfo {author} {\bibfnamefont {P.~D.}\ \bibnamefont {Rack}}, \bibinfo
  {author} {\bibfnamefont {Y.}~\bibnamefont {Guan}}, \ and\ \bibinfo {author}
  {\bibfnamefont {J.}~\bibnamefont {Fowlkes}},\ }\href@noop {} {\bibfield
  {journal} {\bibinfo  {journal} {{Phys. Rev. E}}\ }\textbf {\bibinfo {volume}
  {79}},\ \bibinfo {pages} {026302} (\bibinfo {year} {2009})}\BibitemShut
  {NoStop}%
\bibitem [{\citenamefont {Fowlkes}\ \emph {et~al.}(2011)\citenamefont
  {Fowlkes}, \citenamefont {Kondic}, \citenamefont {Diez},\ and\ \citenamefont
  {Rack}}]{fowlkes_nano11}%
  \BibitemOpen
  \bibfield  {author} {\bibinfo {author} {\bibfnamefont {J.~D.}\ \bibnamefont
  {Fowlkes}}, \bibinfo {author} {\bibfnamefont {L.}~\bibnamefont {Kondic}},
  \bibinfo {author} {\bibfnamefont {J.}~\bibnamefont {Diez}}, \ and\ \bibinfo
  {author} {\bibfnamefont {P.}~\bibnamefont {Rack}},\ }\href@noop {} {\bibfield
   {journal} {\bibinfo  {journal} {Nano Letters}\ }\textbf {\bibinfo {volume}
  {11}},\ \bibinfo {pages} {2478} (\bibinfo {year} {2011})}\BibitemShut
  {NoStop}%
\bibitem [{\citenamefont {Gonzalez}\ \emph
  {et~al.}(2013{\natexlab{a}})\citenamefont {Gonzalez}, \citenamefont {Diez},
  \citenamefont {Wu}, \citenamefont {Fowlkes}, \citenamefont {Rack},\ and\
  \citenamefont {Kondic}}]{lang13}%
  \BibitemOpen
  \bibfield  {author} {\bibinfo {author} {\bibfnamefont {A.}~\bibnamefont
  {Gonzalez}}, \bibinfo {author} {\bibfnamefont {J.}~\bibnamefont {Diez}},
  \bibinfo {author} {\bibfnamefont {Y.}~\bibnamefont {Wu}}, \bibinfo {author}
  {\bibfnamefont {J.}~\bibnamefont {Fowlkes}}, \bibinfo {author} {\bibfnamefont
  {P.}~\bibnamefont {Rack}}, \ and\ \bibinfo {author} {\bibfnamefont
  {L.}~\bibnamefont {Kondic}},\ }\href@noop {} {\bibfield  {journal} {\bibinfo
  {journal} {Langmuir}\ }\textbf {\bibinfo {volume} {13}},\ \bibinfo {pages}
  {9378} (\bibinfo {year} {2013}{\natexlab{a}})}\BibitemShut {NoStop}%
\bibitem [{\citenamefont {Fowlkes}\ \emph {et~al.}(2014)\citenamefont
  {Fowlkes}, \citenamefont {Roberts}, \citenamefont {Wu}, \citenamefont {Diez},
  \citenamefont {Gonz{\'a}lez}, \citenamefont {Hartnett}, \citenamefont
  {Mahady}, \citenamefont {Afkhami}, \citenamefont {Kondic},\ and\
  \citenamefont {Rack}}]{NanoLett14}%
  \BibitemOpen
  \bibfield  {author} {\bibinfo {author} {\bibfnamefont {J.~D.}\ \bibnamefont
  {Fowlkes}}, \bibinfo {author} {\bibfnamefont {N.~A.}\ \bibnamefont
  {Roberts}}, \bibinfo {author} {\bibfnamefont {Y.}~\bibnamefont {Wu}},
  \bibinfo {author} {\bibfnamefont {J.~A.}\ \bibnamefont {Diez}}, \bibinfo
  {author} {\bibfnamefont {A.~G.}\ \bibnamefont {Gonz{\'a}lez}}, \bibinfo
  {author} {\bibfnamefont {C.}~\bibnamefont {Hartnett}}, \bibinfo {author}
  {\bibfnamefont {K.}~\bibnamefont {Mahady}}, \bibinfo {author} {\bibfnamefont
  {S.}~\bibnamefont {Afkhami}}, \bibinfo {author} {\bibfnamefont
  {L.}~\bibnamefont {Kondic}}, \ and\ \bibinfo {author} {\bibfnamefont
  {P.}~\bibnamefont {Rack}},\ }\href {\doibase 10.1021/nl404128d} {\bibfield
  {journal} {\bibinfo  {journal} {Nano Letters}\ }\textbf {\bibinfo {volume}
  {14}},\ \bibinfo {pages} {774} (\bibinfo {year} {2014})}\BibitemShut
  {NoStop}%
\bibitem [{\citenamefont {Mahady}\ \emph {et~al.}(2013)\citenamefont {Mahady},
  \citenamefont {Afkhami}, \citenamefont {Diez},\ and\ \citenamefont
  {Kondic}}]{mahady_13}%
  \BibitemOpen
  \bibfield  {author} {\bibinfo {author} {\bibfnamefont {K.}~\bibnamefont
  {Mahady}}, \bibinfo {author} {\bibfnamefont {S.}~\bibnamefont {Afkhami}},
  \bibinfo {author} {\bibfnamefont {J.}~\bibnamefont {Diez}}, \ and\ \bibinfo
  {author} {\bibfnamefont {L.}~\bibnamefont {Kondic}},\ }\href@noop {}
  {\bibfield  {journal} {\bibinfo  {journal} {Phys. Fluids}\ }\textbf {\bibinfo
  {volume} {25}},\ \bibinfo {pages} {112103} (\bibinfo {year}
  {2013})}\BibitemShut {NoStop}%
\bibitem [{\citenamefont {Diez}\ and\ \citenamefont {Kondic}(2007)}]{dk_pof07}%
  \BibitemOpen
  \bibfield  {author} {\bibinfo {author} {\bibfnamefont {J.}~\bibnamefont
  {Diez}}\ and\ \bibinfo {author} {\bibfnamefont {L.}~\bibnamefont {Kondic}},\
  }\href@noop {} {\bibfield  {journal} {\bibinfo  {journal} {{Phys. Fluids}}\
  }\textbf {\bibinfo {volume} {19}},\ \bibinfo {pages} {072107} (\bibinfo
  {year} {2007})}\BibitemShut {NoStop}%
\bibitem [{\citenamefont {Diez}\ and\ \citenamefont {Kondic}(2002)}]{DK_jcp02}%
  \BibitemOpen
  \bibfield  {author} {\bibinfo {author} {\bibfnamefont {J.}~\bibnamefont
  {Diez}}\ and\ \bibinfo {author} {\bibfnamefont {L.}~\bibnamefont {Kondic}},\
  }\href@noop {} {\bibfield  {journal} {\bibinfo  {journal} {J. Comput. Phys.}\
  }\textbf {\bibinfo {volume} {183}},\ \bibinfo {pages} {274} (\bibinfo {year}
  {2002})}\BibitemShut {NoStop}%
\bibitem [{\citenamefont {Dong}\ and\ \citenamefont {Kondic}(2016)}]{in_prep}%
  \BibitemOpen
  \bibfield  {author} {\bibinfo {author} {\bibfnamefont {N.}~\bibnamefont
  {Dong}}\ and\ \bibinfo {author} {\bibfnamefont {L.}~\bibnamefont {Kondic}},\
  }\href@noop {} {} (\bibinfo {year} {2016}),\ \bibinfo {note} {in
  preparation}\BibitemShut {NoStop}%
\bibitem [{\citenamefont {Lin}\ \emph {et~al.}(2012)\citenamefont {Lin},
  \citenamefont {Kondic},\ and\ \citenamefont {Filippov}}]{lin_pof12}%
  \BibitemOpen
  \bibfield  {author} {\bibinfo {author} {\bibfnamefont {T.-S.}\ \bibnamefont
  {Lin}}, \bibinfo {author} {\bibfnamefont {L.}~\bibnamefont {Kondic}}, \ and\
  \bibinfo {author} {\bibfnamefont {A.}~\bibnamefont {Filippov}},\ }\href@noop
  {} {\bibfield  {journal} {\bibinfo  {journal} {Phys. Fluids}\ }\textbf
  {\bibinfo {volume} {24}},\ \bibinfo {pages} {022105} (\bibinfo {year}
  {2012})}\BibitemShut {NoStop}%
\bibitem [{\citenamefont {Lin}\ \emph {et~al.}(2013)\citenamefont {Lin},
  \citenamefont {Kondic}, \citenamefont {Thiele},\ and\ \citenamefont
  {Cummings}}]{lin_jfm13}%
  \BibitemOpen
  \bibfield  {author} {\bibinfo {author} {\bibfnamefont {T.-S.}\ \bibnamefont
  {Lin}}, \bibinfo {author} {\bibfnamefont {L.}~\bibnamefont {Kondic}},
  \bibinfo {author} {\bibfnamefont {U.}~\bibnamefont {Thiele}}, \ and\ \bibinfo
  {author} {\bibfnamefont {L.~J.}\ \bibnamefont {Cummings}},\ }\href@noop {}
  {\bibfield  {journal} {\bibinfo  {journal} {J. Fluid Mech.}\ }\textbf
  {\bibinfo {volume} {729}},\ \bibinfo {pages} {214} (\bibinfo {year}
  {2013})}\BibitemShut {NoStop}%
\bibitem [{\citenamefont {Gonzalez}\ \emph
  {et~al.}(2013{\natexlab{b}})\citenamefont {Gonzalez}, \citenamefont {Diez},\
  and\ \citenamefont {Kondic}}]{gdk_jfm13}%
  \BibitemOpen
  \bibfield  {author} {\bibinfo {author} {\bibfnamefont {A.~G.}\ \bibnamefont
  {Gonzalez}}, \bibinfo {author} {\bibfnamefont {J.~D.}\ \bibnamefont {Diez}},
  \ and\ \bibinfo {author} {\bibfnamefont {L.}~\bibnamefont {Kondic}},\
  }\href@noop {} {\bibfield  {journal} {\bibinfo  {journal} {J. Fluid Mech.}\
  }\textbf {\bibinfo {volume} {718}},\ \bibinfo {pages} {213} (\bibinfo {year}
  {2013}{\natexlab{b}})}\BibitemShut {NoStop}%
\bibitem [{\citenamefont {Witelski}\ and\ \citenamefont
  {Bowen}(2003)}]{witelski_anm03}%
  \BibitemOpen
  \bibfield  {author} {\bibinfo {author} {\bibfnamefont {T.}~\bibnamefont
  {Witelski}}\ and\ \bibinfo {author} {\bibfnamefont {M.}~\bibnamefont
  {Bowen}},\ }\href@noop {} {\bibfield  {journal} {\bibinfo  {journal} {Applied
  Numer. Math.}\ }\textbf {\bibinfo {volume} {45}},\ \bibinfo {pages} {331}
  (\bibinfo {year} {2003})}\BibitemShut {NoStop}%
\bibitem [{\citenamefont {Nesic}\ \emph {et~al.}(2015)\citenamefont {Nesic},
  \citenamefont {Cuerno}, \citenamefont {Moro},\ and\ \citenamefont
  {Kondic}}]{nesic_pre15}%
  \BibitemOpen
  \bibfield  {author} {\bibinfo {author} {\bibfnamefont {S.}~\bibnamefont
  {Nesic}}, \bibinfo {author} {\bibfnamefont {R.}~\bibnamefont {Cuerno}},
  \bibinfo {author} {\bibfnamefont {E.}~\bibnamefont {Moro}}, \ and\ \bibinfo
  {author} {\bibfnamefont {L.}~\bibnamefont {Kondic}},\ }\href@noop {}
  {\bibfield  {journal} {\bibinfo  {journal} {Phys. Rev. E}\ }\textbf {\bibinfo
  {volume} {92}},\ \bibinfo {pages} {061002(R)} (\bibinfo {year}
  {2015})}\BibitemShut {NoStop}%
\bibitem [{\citenamefont {Diez}\ \emph {et~al.}(2016)\citenamefont {Diez},
  \citenamefont {Gonz\'alez},\ and\ \citenamefont {Fern\'andez}}]{diez_pre16}%
  \BibitemOpen
  \bibfield  {author} {\bibinfo {author} {\bibfnamefont {J.~A.}\ \bibnamefont
  {Diez}}, \bibinfo {author} {\bibfnamefont {A.~G.}\ \bibnamefont
  {Gonz\'alez}}, \ and\ \bibinfo {author} {\bibfnamefont {R.}~\bibnamefont
  {Fern\'andez}},\ }\href@noop {} {\bibfield  {journal} {\bibinfo  {journal}
  {Phys. Rev. E}\ }\textbf {\bibinfo {volume} {93}},\ \bibinfo {pages} {013120}
  (\bibinfo {year} {2016})}\BibitemShut {NoStop}%
\end{thebibliography}%

\end{document}